\documentclass[%
 aip,
 jcp,
 amsmath,
 amssymb,
 reprint
]{revtex4-2}

\usepackage{graphicx}
\usepackage{dcolumn}
\usepackage{bm}

\usepackage[utf8]{inputenc}
\usepackage[T1]{fontenc}
\usepackage{mathptmx}

\usepackage[acronym]{glossaries}
\newacronym{DFT}{DFT}{density functional theory}
\newacronym{TDDFT}{TDDFT}{time-dependent density functional theory}
\newacronym{SnPc}{\textbf{SnPc}}{tin(II) phthalocyanine}
\newacronym{IMDHOM}{IMDHOM}{independent-mode displaced harmonic oscillator model}
\newacronym{LMCT}{LMCT}{ligand-to-metal charge transfer}
\newacronym{MLCT}{MLCT}{metal-to-ligand charge transfer}
\newacronym{MC}{MC}{metal-centered}
\newacronym{SERS}{SERS}{surface-enhanced Raman spectroscopy}
\newacronym{TERS}{TERS}{tip-enhanced Raman spectroscopy}

\usepackage{lmodern}
\usepackage[ngerman,english]{babel}
\usepackage{amsmath}
\usepackage{amsfonts}
\usepackage{amssymb}
\usepackage{nicefrac}
\usepackage{siunitx}
\DeclareSIUnit{\arbu}{arb.u.}

\begin{document}

\title[Are charged tips driving TERS-resolution? A full quantum chemical approach]{Are charged tips driving TERS-resolution? A full quantum chemical approach}

\author{K. Fiederling}
\author{S. Kupfer}
\email{stephan.kupfer@uni-jena.de}
\author{S. Gräfe}
\email{s.graefe@uni-jena.de}
\affiliation{Institut für Physikalische Chemie and Abbe Center of Photonics, Friedrich-Schiller-Universität Jena, Helmholtzweg 4, 07743 Jena, Germany}

\date{\today}

\begin{abstract}

Experimental evidence suggests an extremely high, possibly even sub-molecular, spatial resolution of \acrfull{TERS}. 
While the underlying mechanism is currently still under discussion, two main contributions are considered: 
The involved plasmonic particles are able to highly confine light to small spatial regions in the near-field, i.e. the electromagnetic effect, 
and the chemical effect due to altered molecular properties of the sample in close proximity to the plasmonic tip. 
Significant theoretical effort is put into the modelling of the electromagnetic contribution by various groups. 
In contrast, we previously introduced a computational protocol that allows for investigation of the local chemical effect -- including non-resonant, resonant and charge transfer contributions -- in a plasmonic hybrid system by mapping the sample molecule with a metallic tip model at the (time-dependent) density functional level of theory. 
In the present contribution, we evaluate the impact of static charges localized on the tip's frontmost atom, possibly induced by the tip geometry in the vicinity of the apex, on the \acrshort{TERS} signal and the lateral resolution. 
To this aim, an immobilized molecule, i.e. \acrfull{SnPc}, is mapped by the plasmonic tip modeled by a single positively vs. negatively charged silver atom. 
The performed quantum chemical simulations reveal a pronounced enhancement of the Raman intensity under non-resonant and resonant conditions with respect to the uncharged reference system, while the contribution of charge transfer phenomena as well as of locally excited states of \acrshort{SnPc} is highly dependent on the tip's charge.

\end{abstract}

\maketitle

\section{Introduction}

The ultimate goal of microscopy is to make the smallest things visible, and what could be smaller than the building blocks of matter: atoms and molecules.
To achieve this goal, resolution in the Angstrom scale is necessary -- but Abbe's diffraction limit is, for the visible spectrum of light, several orders of magnitude above this scale and therefore makes this seemingly impossible.
However, some modern techniques are able to circumvent this limit through different means, like 
fluorescence-based stimulated emission depletion (STED),\cite{Hell_Opt.Lett.OL_1994} 
photoactivated localization microscopy (PALM), \cite{Betzig_Science_2006}
stochastic optical reconstruction microscopy (STORM), \cite{Rust_Nat.Methods_2006}
and near-field methods like scanning near-field optical microscopy (SNOM),\cite{Ash_Nature_1972}
and tip-enhanced Raman spectroscopy (TERS).\cite{Anderson_Appl.Phys.Lett._2000,Hayazawa_OpticsCommunications_2000,Stockle_ChemicalPhysicsLetters_2000}

The mechanism behing TERS (and the closely related but not resolution-driven surface-enhanced Raman spectroscopy, SERS\cite{Fleischmann_ChemicalPhysicsLetters_1974,Albrecht_J.Am.Chem.Soc._1977,Jeanmaire_JournalofElectroanalyticalChemistryandInterfacialElectrochemistry_1977,Langer_ACSNano_2020}) 
is based on surface plasmons, leading to an increase in Raman signal by up to six orders of magnitude.\cite{Jiang_NanoLett._2012,Pettinger_Phys.Rev.Lett._2004}
This is commonly attributed to two different effects:
The electromagnetic effect\cite{Corni_ChemicalPhysicsLetters_2001, Corni_J.Chem.Phys._2001a, Chulhai_J.Phys.Chem.C_2013, Chiang_NanoLett._2015, Zhang_J.Phys.Chem.C_2015, Morton_Chem.Rev._2011, Thomas_J.RamanSpectrosc._2013}
strongly increases the Raman signal of a sample close to the plasmonic particle by enhancing and confining the local electric field, 
especially in a so-called pico-cavity between the plasmonic tip and a metal substrate. \cite{Zhang_Phys.Rev.B_2014, Benz_Science_2016, Trautmann_Nanoscale_2016, Chen_Nanoscale_2018, Urbieta_ACSNano_2018}
Furthermore, the field-induced dipole in the sample molecule is accompanied by a mirror dipole in the plasmonic particle, modifying and further enhancing the near-field.\cite{Atay_NanoLett._2004,Morton_J.Chem.Phys._2010,Morton_J.Chem.Phys._2011}
On the other hand, the chemical effect \cite{Zhao_J.Am.Chem.Soc._2006, Jensen_J.Phys.Chem.C_2007, Jensen_Chem.Soc.Rev._2008, Liu_SpectrochimicaActaPartA:MolecularandBiomolecularSpectroscopy_2009, Valley_J.Phys.Chem.Lett._2013, Latorre_Phys.Chem.Chem.Phys._2015, Morton_Chem.Rev._2011}
describes the close-range interactions between the sample molecule and the plasmonic particle.
Scanning the tip over the sample molecule leads to different site-specific tip-sample interaction and therefore can shift position and intensity of Raman bands, with extreme sensitivity to the local chemical environment.\cite{Thomas_J.RamanSpectrosc._2013, Latorre_Nanoscale_2016}
The chemical effect is typically further divided into three parts:
Ground-state contributions from non-resonant chemical  interactions between tip and sample,
resonant contributions due to excited states of the molecule with similar energy as the excitation radiation,
and charge transfer phenomena between tip and sample.\cite{Jensen_Chem.Soc.Rev._2008, Sun_Phys.Chem.Chem.Phys._2009, Avila_Chem.Commun._2011a, Avila_Chem.Commun._2011, Morton_Chem.Rev._2011,Xia_J.RamanSpectrosc._2014}

Recent experiments under cryogenic and even ambient conditions hint at possible lateral resolutions of TERS of \SI{1}{\nano\meter} or even below \cite{Zhang_Nature_2013,Fang_Phys.Chem.Chem.Phys._2014,Klingsporn_J.Am.Chem.Soc._2014,Sun_Adv.Opt.Mater._2014,He_J.Am.Chem.Soc._2019,Jiang_Nat.Nanotechnol._2015,Lee_Nature_2019,Richard-Lacroix_Chem.Soc.Rev._2017}
and therefore providing up to (sub-)molecular resolution.
At first this is surprising, considering that the employed plasmonic nanoparticles typically have a diameter of 20 to \SI{40}{\nano\meter}.\cite{Martin_JournalofAppliedPhysics_2001,Geshev_Phys.Rev.B_2004,Notingher_J.Phys.Chem.B_2005,Payton_Acc.Chem.Res._2014}
Evidence suggests that small, atomic-sized features on the plasmonic particle are the source of this resolution \cite{Trautmann_Nanoscale_2016,Chen_Nanoscale_2018,Barbry_NanoLett._2015,Schmidt_ACSNano_2016}
and deeper atomic layers in the particle play only a minor role. \cite{Latorre_Nanoscale_2016,Trautmann_Nanoscale_2018}

There are several approaches to model these experiments, for example developed in the groups of Schatz,\cite{Hao_J.Chem.Phys._2003,Zou_J.Chem.Phys._2004,Gieseking_J.Phys.Chem.A_2016,Ding_J.Phys.Chem.C_2018}
Jensen,\cite{Payton_J.Chem.Phys._2012,Payton_Acc.Chem.Res._2014,Hu_J.Chem.TheoryComput._2016,Liu_ACSNano_2017,Chen_Nanoscale_2018,Chen_Nat.Commun._2019}
and Aizpurua.\cite{Zhang_Nature_2013,Barbry_NanoLett._2015,Benz_Science_2016,Schmidt_ACSNano_2016,Langer_ACSNano_2020}
The discrete interaction model/quantum mechanical (DIM/QM) method\cite{Morton_J.Chem.Phys._2010,Morton_J.Chem.Phys._2011}
by Jensen et al. provides a combination of \acrfull{TDDFT} for the sample molecule with atomistic electrodynamics simulation for the metallic particle.
Aizpurua et al. and Sánchez-Portal et al. treat the plasmonic cavity on a full quantum electrodynamical level and thereby show that small, atomic-sized features on the nanoparticle can confine and enhance electromagnetic radiation into small enough volumes to provide sub-nanometer resolution.\cite{Barbry_NanoLett._2015,Schmidt_ACSNano_2016}
Furthermore, by employing an effective Hamiltonian containing the highly confined electric field of a plasmonic particle, Luo et al. describe the tip-sample interaction quantum mechanically.\cite{Duan_J.Am.Chem.Soc._2015,Duan_Angew.Chem.Int.Ed._2016,Xie_Nanoscale_2018}

In another different approach, we introduced a computational protocol to describe highly location-specific tip-sample interactions entirely quantum chemically,\cite{Latorre_Nanoscale_2016}
in essence covering the ground state contribution to the chemical effect.
This model was expanded later on to also include excited-state and charge transfer phenomena by including the Raman response to light in resonance with molecular excited states.\cite{Fiederling_Nanoscale_2020}
To achieve this,  the immobilized sample molecule is three-dimensionally mapped by placing a model of the plasmonic particle on different positions relative to the molecule.
By constraining the tip model to a small cluster or even a single atom, this allows us to describe the whole hybrid system at a fairly high \acrfull{DFT} level and gives insight into the changes to molecular properties occuring by moving the tip in every possible spatial direction.
Minute movements of the tip can result in large changes of spectral position and, especially, intensity of Raman bands, explaining the sub-molecular resolution.
Excitation at resonant wavelengths on the other hand, provides a means to enhance the intensity of certain vibrational modes by several orders of magnitude, depending on the respective excited states in resonance.

\begin{figure}
\includegraphics{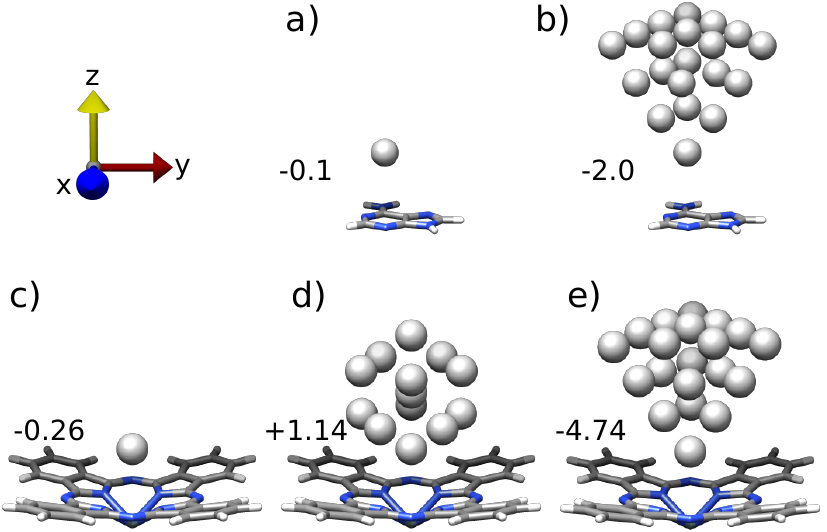}
\caption{\label{fig:tips} Mulliken charge of the apex tip atom for different tip models near adenine (a) single Ag atom, b) tetrahedral ($\mathrm{Ag}_{20}$) cluster; from Ref.~\onlinecite{Latorre_Nanoscale_2016}); or \textbf{SnPc}: c) single Ag atom, d) icosahedral ($\mathrm{Ag}_{13}$), e) tetrahedral ($\mathrm{Ag}_{20}$) cluster.}
\end{figure}

In the current contribution, we elaborate on the influence of the charge of the frontmost (silver) tip atom on the ground and excited state properties of such plasmonic hybrid systems,
and finally its impact on the local Raman response and the lateral resolution.
This study is motivated by the observed variation of Raman signals and intensity maps of adenine mapped either by a single silver atom or a small tetrahedral cluster comprised of 20 silver atoms, see Fig.~\ref{fig:tips}a) and b).\cite{Latorre_Nanoscale_2016} 
An analysis of both neutral hybrid systems suggests that the polarizability is highly dependent on the charge of the Ag atom chemically interacting with the sample.
Previous simulations show merely a small negative charge of \SI{-0.1}{\elementarycharge} for the single-atom tip model.
However, the silver cluster model features also a total charge of \SI{-0.1}{\elementarycharge} in the hybrid system; its four vertices feature considerable negative partial charges, in particular -- with \SI{-2.0}{\elementarycharge} -- the vertex chemically interacting with the sample.
In addition, preliminary \acrshort{DFT} results on various clusters models, i.e. an icosahedron ($\mathrm{Ag}_{13}$) vs. a tetrahedron ($\mathrm{Ag}_{20}$) shown in Fig.~\ref{fig:tips}d) and e), suggest that positive as well as negative charges of the frontmost tip atom are present depending on the tip geometry -- 
possibly influencing the chemical interaction and, thus, the \acrshort{TERS} signal.
Thus, the shape of the metallic nanoparticle might yield a specific chemical interaction in plasmon-enhanced spectroscopy.
However, the interest in charged tips is not necessarily limited to theoretical investigations addressing structurally induced charge localization effects of the Raman signal, but charged tips may be introduced purposely in the experimental setup by applying an external electric potential.
This way, the molecule's Raman signal can be obtained in the cavity of the tip and the immobilizing surface as a function of the applied voltage -- resulting in (partially) charged tip atoms.\cite{Lee_Sci.Adv._2018,Peller_Nat.Photonics_2020}
The chosen model system is similar to our previously studied one, consisting of an immobilized \acrfull{SnPc} molecule that is mapped by a single, positively or negatively charged silver atom on the \acrshort{DFT} and \acrshort{TDDFT} level.
In particular, we investigate singly positively vs. singly negatively charged silver tips in comparison to the uncharged reference.
Planar molecules, such as (metallo-)porphyrins\cite{Zhang_Nature_2013,Meng_Sci.Rep._2015,Chiang_NanoLett._2016,Liu_ACSNano_2017,Lee_Nature_2019} 
and phthalocyanines\cite{Stadler_Nat.Phys._2009,Kroger_NewJ.Phys._2010,Birmingham_J.Phys.Chem.C_2018,Jaculbia_Nat.Nanotechnol._2020,Jaculbia_ApplSpectrosc_2020} 
have been the subject of several studies in the context of plasmon-enhanced spectroscopy like SERS and TERS.
\textbf{SnPc} in particular features bright excited states that can respond to light at different wavelengths and allows for possible charge transfer to or from the tip, both in the visible spectrum.

\section{Computational Details}

To simulate the \acrfull{TERS} experiment, the same protocol as previously published by our group\cite{Latorre_Nanoscale_2016,Fiederling_Nanoscale_2020}  was employed.
In that, an immobilized molecule, i.e \acrfull{SnPc}, is mapped by a silver tip, in this case modelled by a single silver atom, along a three-dimensional grid.

\begin{figure}
\includegraphics{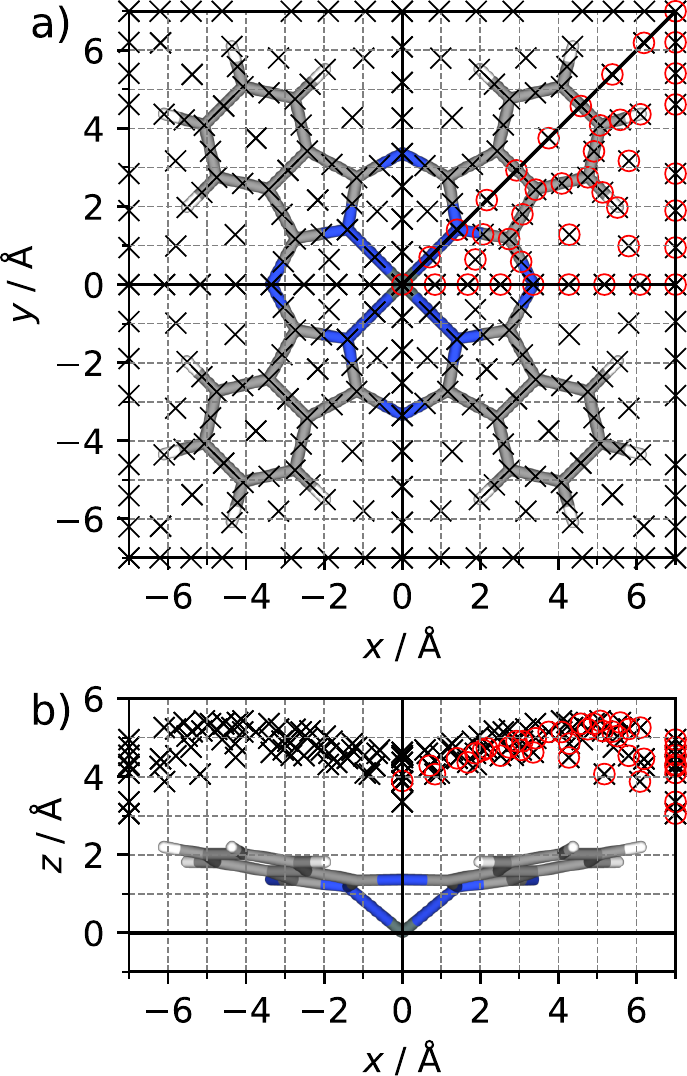}
\caption{\label{fig:setup} All 253 considered tip positions (black crosses) over the \acrshort{SnPc} molecule and the 41 explicitly calculated points (in red circles), as introduced previously.\cite{Fiederling_Nanoscale_2020} 
a) View along $z$-axis b) View along $y$-axis.}
\end{figure}

The geometry of the sample molecule, \acrshort{SnPc}, was optimized and a vibrational analysis ensured that a minimum on the $3N-6$ dimensional potential energy hypersurface has been obtained.
The molecule was then oriented parallel to the $x,y$-plane, with the Sn atom localized in the origin of coordinates, and a single silver atom was placed on different 3D-positions above the sample, see Fig.~\ref{fig:setup}.
The $z$-coordinate (height) of the silver atom was chosen for each tip-position based on the van-der-Waals radii of \acrshort{SnPc} and the silver atom; identical grids were applied for the charged hybrid systems.
As shown previously by our group,\cite{Fiederling_Nanoscale_2020} such van-der-Waals mapping allows to assess the local bonding region between the silver tip and the sample in reasonable agreement with computationally more demanding partial relaxations
($x,y$ coordinates frozen) within each position using \acrfull{DFT} calculations.
Using the $C_\mathrm{4v}$ symmetry of the \acrshort{SnPc} molecule, only 41 unique positions had to be considered explicitly, while the whole molecular map was obtained upon applying symmetry considerations.
Additionally, the influence of the ``height'' ($z$-coordinate) of the silver tip above the molecule was studied exemplarily for two grid positions (see Fig.~\ref{fig:pes} and Fig.~S5) in the range of 2.5 to \SI{20}{\angstrom} for the uncharged as well as for the charged tip mimics.

As a result of the unpaired electron in the silver atom (ground state electron configuration: $5s^1 4d^{10}$), the uncharged system is of doublet multiplicity with the spin density mainly residing in the $5s$ orbital of the silver atom; this hybrid system is denoted \textbf{SnPc-Ag}.
For the charged hybrid systems, the multiplicity was restricted to a singlet which localises the charge mainly on the silver atom.
These systems, featuring a positively or negatively charged tip, are labeled \textbf{SnPc-Ag\textsuperscript{+}} and \textbf{SnPc-Ag\textsuperscript{-}}, respectively.

To assess the possible charge-induced lateral resolution limit of TERS -- based on the chemical interaction among \textbf{SnPc} and the (charged) tip -- 3D-grid simulations were performed, as outlined before.
Therefore, on each position, the vibrational modes $q_l$ including their polarizability derivatives $\nicefrac{\partial \alpha}{\partial q_l}$ were calculated, as well as vertical excitation energies $E_e$, excited-state gradients $\nicefrac{\partial E_e}{\partial q_l}$ and transition dipole moments $\mu_e$ for all relevant excited states $e$.

Contributions of all excited states within the excitation energy window of $0 < E_e \le \SI{3.2}{\electronvolt}$ -- corresponding to an excitation  in the visible region -- and with an oscillator strength of $f \ge 0.001$ were considered.
Normal modes with a wavenumber of $\tilde{\nu}_l < \SI{100}{\per\centi\meter}$  were not taken into account.

Based on these properties, the \acrfull{IMDHOM} was employed. 
Here, it is assumed that the electronic ground- and excite- state potentials are harmonic and merely displaced in the equilibrium position and, hence, share the same set of vibrational modes. 
This allows calculating the $zz$-component of the transition polarizabilities as: \cite{Guthmuller_J.Chem.Phys._2016}
\begin{equation}\label{eq:alpha}
\left(\alpha_{zz}\right)_{g0_l \rightarrow g1_l} = \sum_e \left(\mu_z \right)_e^2 \frac{\Delta_{e,l}}{\sqrt{2}} \left( \Phi_e (E_\mathrm{L}) - \Phi_e (E_\mathrm{L} - E_l )\right) \;.
\end{equation}
With the dimensionless displacement $\Delta_{e,l}$, defined as:
\begin{equation}
 \Delta_{e,l} = - \frac{\hbar^2}{\sqrt{E_l^3}} \left( \frac{\partial E_e}{\partial q_l} \right)_0 \;,
\end{equation}
and the function $\Phi_e$, neglecting Franck-Condon factors, given by:
\begin{equation}
 \Phi_e (E_\mathrm{L}) = \frac{1}{E_{e,g} - E_\mathrm{L} -\mathrm{i} \Gamma} \;.
\end{equation}
In the equations above, $E_{e,g}$ is the vertical excitation energy from the ground-state $g$ to excited-state $e$ and $\Gamma$ a damping factor describing homogeneous broadening (chosen as $\SI{3000}{\per\centi\meter} \widehat{=} \SI{0.372}{\electronvolt}$).
Detailed information with respect to the computational methods has been previously reported.\cite{Fiederling_Nanoscale_2020,Guthmuller_J.Chem.Phys._2016,Wachtler_CoordinationChemistryReviews_2012}

Assuming the detection of the Raman signal occurs mostly in $z$-direction (the direction of the metal tip), the intensity $I_l$ of each mode is determined by the $zz$-component of the respective derivative of the (transition) polarizability tensor
\begin{equation}
I_l = \left( E_\mathrm{L} - E_l \right) ^4 \left| \frac{\partial \left( \alpha_{zz} \right)_{g0_l \rightarrow g1_l}}{\partial q_l} \right| ^2.
\end{equation}

All calculations were performed on the \acrshort{DFT} and \acrshort{TDDFT} levels of theory using the Gaussian~16 program\cite{g16} with the range-separated CAM-B3LYP functional\cite{Yanai_ChemicalPhysicsLetters_2004} and the 6-311+G** triple-$\zeta$ basis set.\cite{Krishnan_J.Chem.Phys._1980, Clark_J.Comput.Chem._1983} 
The metal atoms (tin, silver) were described with the electronic core potentials MWB46 and MWB28 and their respective basis sets\cite{Andrae_Theoret.Chim.Acta_1990}, respectively.
The D3 dispersion correction with Becke-Johnson dampening was employed for all calculations.\cite{Grimme_J.Comput.Chem._2011}

\section{Results}

This section elaborates on the simulated \acrshort{TERS} response of both charged hybrid systems (i.e. \textbf{SnPc-Ag\textsuperscript{+}} and \textbf{SnPc-Ag\textsuperscript{-}}) at the above mentioned level of theory, in comparison to the uncharged hybrid system (\textbf{SnPc-Ag}). 
First, the non-resonant case is discussed, observed for incident radiation with energy far away from any electronic excitations of the system, by employing exemplarily a \SI{1064}{\nano\meter} laser.
In the following, two laser wavelengths in the visible spectrum (\SI{633}{\nano\meter} and \SI{442}{\nano\meter}) are chosen to probe the excited-state landscape and investigate different resonant Raman conditions.
For a more detailed discussion of the uncharged \textbf{SnPc-Ag} system obtained using a similar computational protocol, see Ref.~\onlinecite{Fiederling_Nanoscale_2020}.

In part, results are presented in the form of intensity maps, where either the (resonant) \acrshort{TERS} intensity of a certain mode is shown for different tip positions above the sample molecule, or the sum over the intensities of all modes.
While the former gives more detailed insight which parts of the molecule respond in a certain, very narrow wavenumber region, the latter provides an overview over the general ability of the system to provide strong \acrshort{TERS} responses for the different tip positions.

\subsection{Non-resonant}

\begin{figure*}
\includegraphics{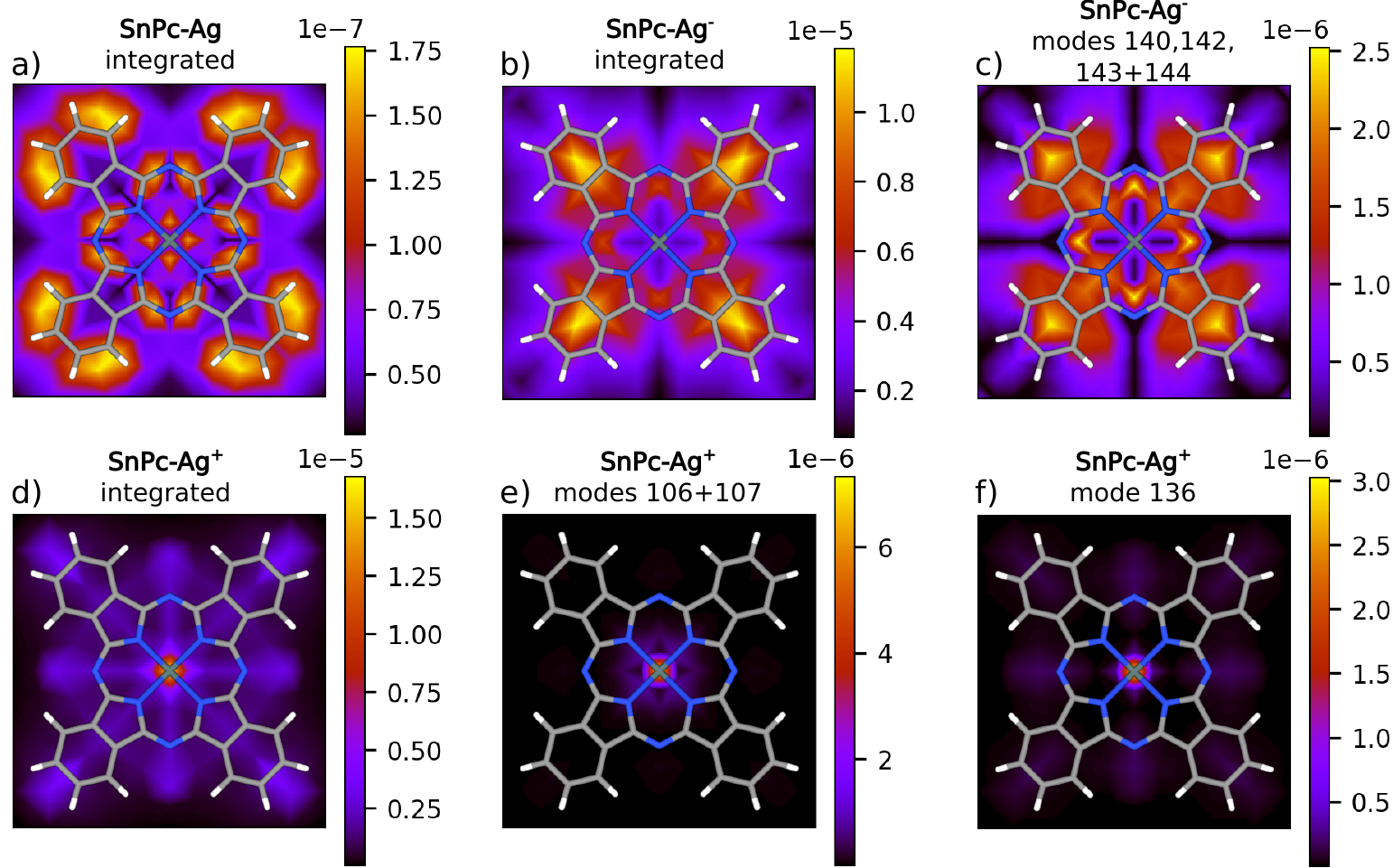}
\caption{\label{fig:non-res}a) Integrated intensity map of the uncharged \textbf{SnPc-Ag} hybrid system 
b) integrated intensity map and
c) map of modes 140, 142, 143, and 144 in the negatively charged hybrid system \textbf{SnPc-Ag\textsuperscript{-}}
d) integrated intensity map,
e) map of modes 106 and 107, and
f) map of mode 136 in \textbf{SnPc-Ag\textsuperscript{+}}.}
\end{figure*}

Starting with the uncharged hybrid system, \textbf{SnPc-Ag}, there are three areas of high intensity that roughly form three concentric rings over the sample, as shown in Fig.~\ref{fig:non-res}a):
The structure in the center of the molecule stems mostly from two low-wavenumber modes involving stretching of  the inner C-N bonds (mode 90) as well as an out-of-plane vibration involving the whole molecule (mode 54).
At this tip position, both modes have roughly the same intensity with \num{2.4e-8} and \SI{2.1e-8}{\arbu} respectively, and combined, they contribute to about a quarter of the overall intensity (\SI{1.6e-7}{\arbu}).
The other two ring-like structures are mainly attributed to C-H in-plane modes and stretching of the outer C-N bonds.
The middle structure follows the inner C-N ring of the molecule and heavily features a C-N-stretching vibration (mode 140) that alone contributes about a sixth (\SI{3.0e-7}{\arbu}) to the overall intensity to the map.
Other notable vibrations in this region (in order of decreasing intensity) stem from modes 133, 115, 116, and 103 (all C-H in-plane) with \num{2.2e-7},\num{ 2.2e-7}, \num{1.8e-7}, and \SI{1.8e-7}{\arbu} intensity, respectively, that combine to nearly half of this structure's brightness.
Around the C-H bonds at the molecule's terminal, the third bright spot again originates from mode 140, but this time with less intensity (\SI{1.8e-7}{\arbu}).
Additionally, modes 149 and 96, representing a ring deformation and a C-H stretching vibration, have their intensity maxima in this region and add to this relatively large region of high intensity.

In contrast, upon introduction of a charge at the tip, i.e. in case of \textbf{SnPc-Ag\textsuperscript{+}} and \textbf{SnPc-Ag\textsuperscript{-}}, the contribution of C-H vibrations to the overall Raman signal generally decreases in favor of modes involving C-N bonds.

For the positively charged tip (\textbf{SnPc-Ag\textsuperscript{+}}), this means that slightly lower wavenumber vibrations, especially those involving whole pyrrole rings, dominate the intensity map.
The bright central spot in Fig.~\ref{fig:non-res}d) is mainly due to mode 107 [Fig.~\ref{fig:non-res}e), together with mode 106, see below] which alone contributes \SI{5.3e-6}{\arbu}, around a third of the overall intensity on this grid point.
Modes 136, see Fig.~\ref{fig:non-res}f), and 106 (shown together with mode 107, as they are very close in wavenumbers) as the two next brightest ones with intensities of \num{3.1e-6} and \SI{2.3e-6}{\arbu} respectively together make up the second third of summed up intensity.
All three modes involve vibrations of C-N bonds: mode 107 mainly involves the nitrogen atoms coordinating the central Sn atom, mode 136 in contrast distorts the C-N bonds connecting the isoindole moieties, and mode 106 is a stretching vibration of the pyrrole rings.

While in case of a positively charged tip, \textbf{SnPc-Ag\textsuperscript{+}}, slightly lower wavenumbers modes, i.e. between 1150 and \SI{1180}{\per\centi\meter}, contribute most significantly, higher wavenumber modes (1550 to \SI{1660}{\per\centi\meter}) are most prominent for \textbf{SnPc-Ag\textsuperscript{-}}.
The intensity map in Fig.~\ref{fig:non-res}b) shows a broader intensity distribution mainly over the isoindole mioeties and the bridging N atoms between them.
Even though this broad distribution involves several tip positions, only very few closely clustered modes make up the majority of the overall intensity, and all of them feature vibrations of the bridging C-N bonds.
Mode 140 is the most intense one with \SI{2.2e-6}{\arbu}, followed by mode 142 (\SI{1.7e-6}{\arbu}), mode 144 (\SI{1.5e-6}{\arbu}), and mode 143 (\SI{1.4e-6}{\arbu}); since their wavenumbers are very close, their summed intensity is shown in Fig.~\ref{fig:non-res}c).

As evident from the integrated intensity maps, illustrated in Fig.~\ref{fig:non-res}a) vs. b) and d), over all modes, the overall intensity of the system's non-resonant \acrshort{TERS} response increases by roughly two orders of magnitude after the introduction of a charge on the tip atom.
This change is not only due to an intensity change in the few modes presented above, but rather drastically changes the whole spectrum of the \textbf{SnPc-Ag} hybrid system.

\begin{figure*}
 \includegraphics[scale=0.65]{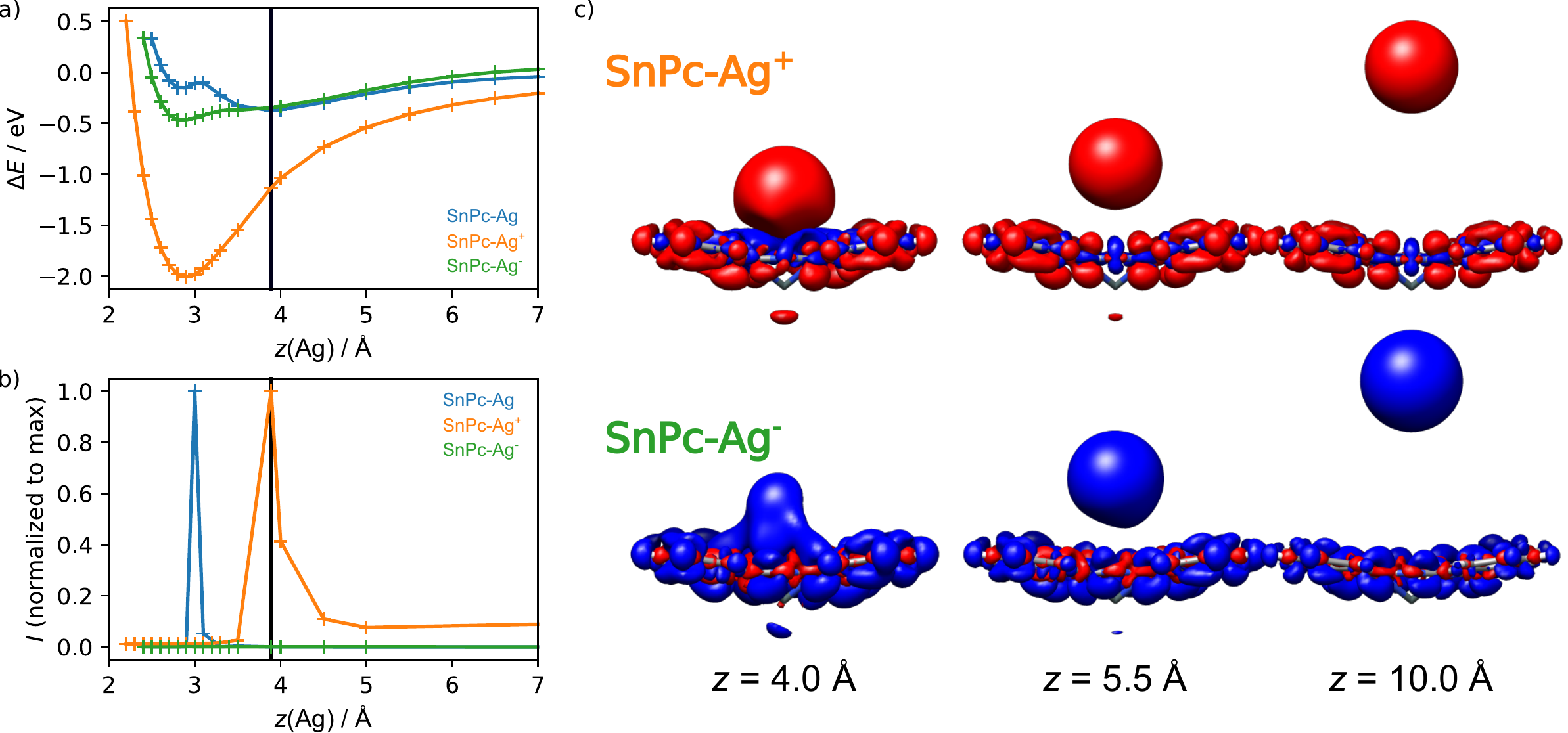}
 \caption{\label{fig:pes} (a) Electric ground state potential energy curves (\textbf{SnPc-Ag\textsuperscript{+}} and \textbf{SnPc-Ag\textsuperscript{-}}: singlet, \textbf{SnPc-Ag}: doublet), (b) normalized TERS intensities and (c) density differences ($\rho_\mathrm{charged} - \rho_\mathrm{neutral}$) for selected $z$-positions, i.e. at $z$~=~10, 5.5 and 4.0~\AA, of the Ag tip mimic at the central position. The vertical black lines in (a) and (b) indicate the bonding region as approximated by the van-der-Waals radii.}
\end{figure*}

In case of the uncharged system, \textbf{SnPc-Ag}, the tip-sample distance -- based on van-der-Waals radii -- is close to the minimum in the potential energy curve for a given $xy$-position of the tip mimic,\cite{Fiederling_Nanoscale_2020} see Fig.~\ref{fig:pes}a) and Fig.~S5 exemplarily.
Notably, the potential energy curve of \textbf{SnPc-Ag} features two minima, the global minimum at roughly 3.9{~\AA} results from the interacting of the silver atom with the central nitrogen atoms coordinating the tin, while the minimum at below 2~{\AA} reveal the Ag-Sn interaction.
However, with the introduced positive or negative charge, the bonding distance is altered for the respective hybrid systems.
 I.e., for both charged systems, the global minimum of the potential energy curve is shifted by roughly \SI{1}{\angstrom} to shorter distances, given by the Ag-Sn interaction, while in case of \textbf{SnPc-Ag\textsuperscript{-}} the N-Ag minimum at 3.9~{\AA} is still observed.
 For \textbf{SnPc-Ag\textsuperscript{-}}, the binding energy between the molecule and the tip is comparable to the one of the uncharged system (0.37 and \SI{0.47}{\electronvolt}, respectively) and the potential energy curve roughly follows the one from the uncharged system for greater tip-sample distances.
 However, the binding energy for the \textbf{SnPc-Ag\textsuperscript{+}} hybrid system is more than four times as large (\SI{2.00}{\electronvolt}) and therefore the curve is steeper, even for larger distances.
 For the TERS intensities shown in Fig.~\ref{fig:pes}b), a roughly exponential increase was expected for smaller tip-sample separations.
 While for \textbf{SnPc-Ag\textsuperscript{+}} and \textbf{SnPc-Ag}, this behavior can be seen (except for very small distances), \textbf{SnPc-Ag\textsuperscript{-}} shows the opposite and decreases the intensity for small distances.
 A partial charge transfer in the ground state, as visible from the density differences of charged and uncharged systems in Fig.~\ref{fig:pes}c), may explain the quenching of TERS intensity for small distances to a certain level.

\subsection{Resonant}

\begin{figure*}
 \includegraphics{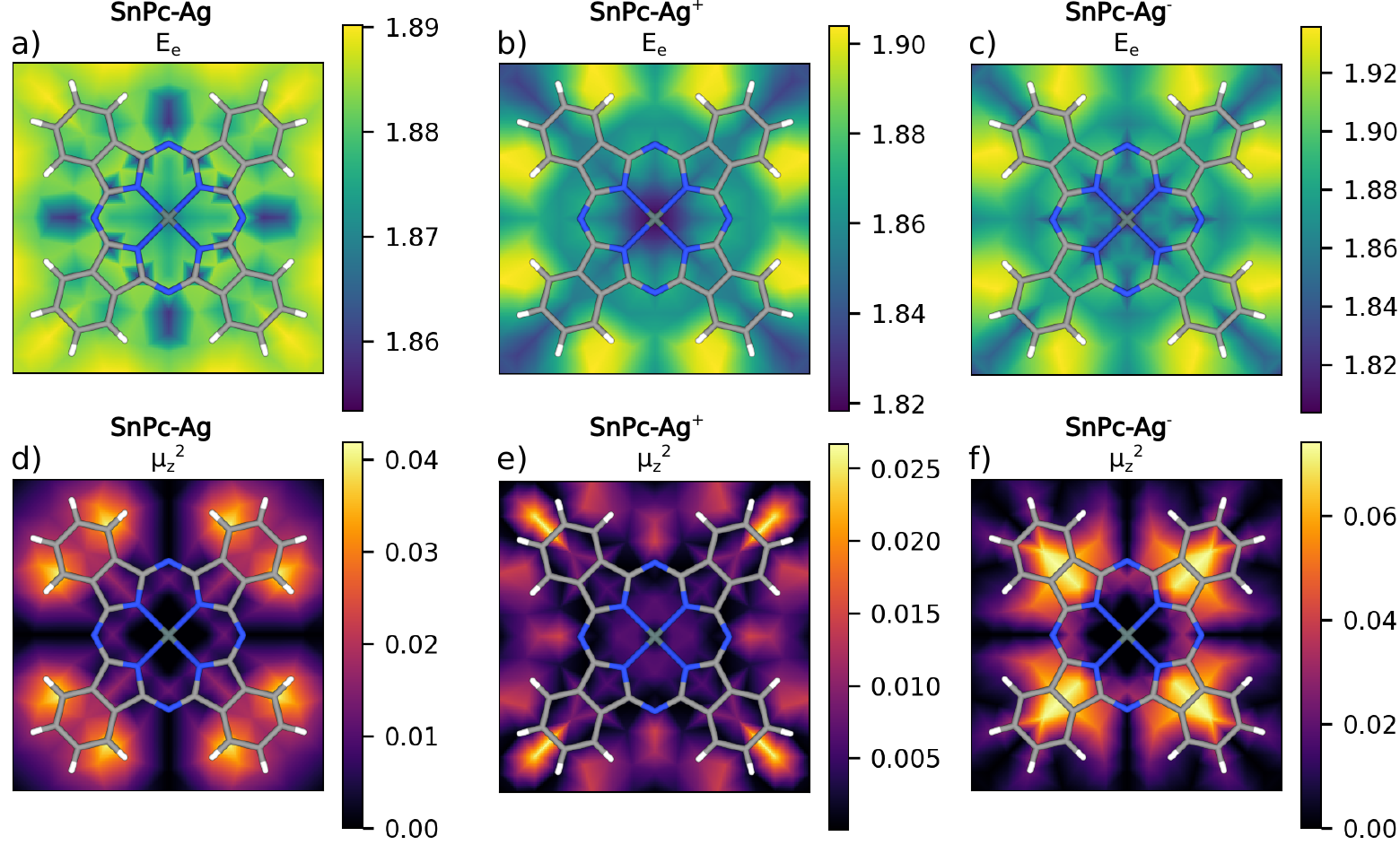}
 \caption{\label{fig:pi} Excitation energies (a)-c); in \si{\electronvolt}) and squared transition dipole moments in $z$-direction $\mu_z^2$ (d)-e); in a.u.) for the first $\pi\pi^*$ transition for different tip positions. 
a)+d) \textbf{SnPc-Ag} 
b)+e) \textbf{SnPc-Ag\textsuperscript{+}} 
c)+f) \textbf{SnPc-Ag\textsuperscript{-}}.} 
\end{figure*}

As previously shown for the uncharged hybrid system, $\pi\pi^*$ transitions of \acrshort{SnPc} are mainly $x,y$-polarized, i.e. in the molecular plane, and thus contribute only minor to the \acrshort{TERS} signal observed in $z$-direction. 
Fig.~\ref{fig:pi} exemplarily shows for the lowest $\pi\pi^*$ state that these states  are relatively insensitive to the position of the silver atom, both in terms of excitation energy (Fig.~\ref{fig:pi}a)) and transition dipole moment (Fig.~\ref{fig:pi}d)), recall Eq.~\ref{eq:alpha}.
This also applies to \textbf{SnPc-Ag\textsuperscript{+}} (Fig.~\ref{fig:pi}b) and e)) and \textbf{SnPc-Ag\textsuperscript{-}} (Fig.~\ref{fig:pi}c) and f)), where these $\pi\pi^*$ states are barely affected neither by the positioning of the tip nor by its charge (see Fig.~S1 and S2).

However, the charge transfer states -- responsible for most of the Raman signal's contrast -- respond strongly to the additional electron or hole on the silver atom.
\textbf{SnPc-Ag\textsuperscript{+}} highly favors electron transfer from the molecule to the positively charged tip and thereby increases the number of low-lying \acrfull{LMCT} states with excitation energies below \SI{3.5}{\electronvolt} 
from merely one in case of \textbf{SnPc-Ag} to about twelve (depending on the tip position) in \textbf{SnPc-Ag\textsuperscript{+}} (see Fig.~S3 and S4).
On the other hand, no \acrfull{MLCT} states could be identified in the same energy region.

The opposite is observed for \textbf{SnPc-Ag\textsuperscript{-}}: 
No \acrshort{LMCT} states below \SI{3.5}{\electronvolt} are predicted by \acrshort{TDDFT}, while the number of \acrshort{MLCT} states is increased from merely two (\textbf{SnPc-Ag}) to approximately 13 -- depending on the tip's position.
This behaviour was to be expected, as electron transfer to an already negatively charged atom is highly unfavourable due to Coulomb repulsion.

The lowest lying charge transfer states of \textbf{SnPc-Ag\textsuperscript{+}} and \textbf{SnPc-Ag\textsuperscript{-}} are (almost) degenerate with the electronic ground state.
Thus, a charge transfer might even be possible within the electronic ground state, altering the character of the ground state upon chemical interaction of sample molecule and tip under (non-resonant) \acrshort{TERS} conditions.

Furthermore, a couple of previously, i.e. in case \textbf{SnPc-Ag}, unencountered \acrfull{MC} transitions of the silver tip are predicted to contribute to the UV/vis spectra of the charged hybrid systems.
For silver nanoparticles, such states are related to plasmonic states and contribute significantly to the \acrshort{TERS} effect.
However, excitation energies and transition dipole moments of such \acrshort{MC} states are expected to be highly sensitive to the size as well as to the structure of the plasmonic nanoparticle.
Thus, the contribution of silver \acrshort{MC} states to the \acrshort{TERS} effect are yet to be described correctly by our  computational protocol.
However, mapping sample molecules, e.g. \acrshort{SnPc}, at the \acrshort{TDDFT} level of theory with realistic metal clusters in a full quantum chemical fashion, which allows an unbiased describtion of the chemical effect, is currently computationally not feasable. 
Therefore, we omit a further discussion of these silver \acrshort{MC} states in the present contribution.

In contrast to the non-resonant Raman effect, the wavelength of the incident light is crucial for resonance Raman since it determines which excited states contribute to the resonance enhancement.
In this contribution, two irradiation wavelengths, namely \SI{633}{\nano\meter} (\SI{1.96}{\electronvolt}) and \SI{442}{\nano\meter} (\SI{2.81}{\electronvolt}), are studied exemplarily.

\begin{figure*}
 \includegraphics{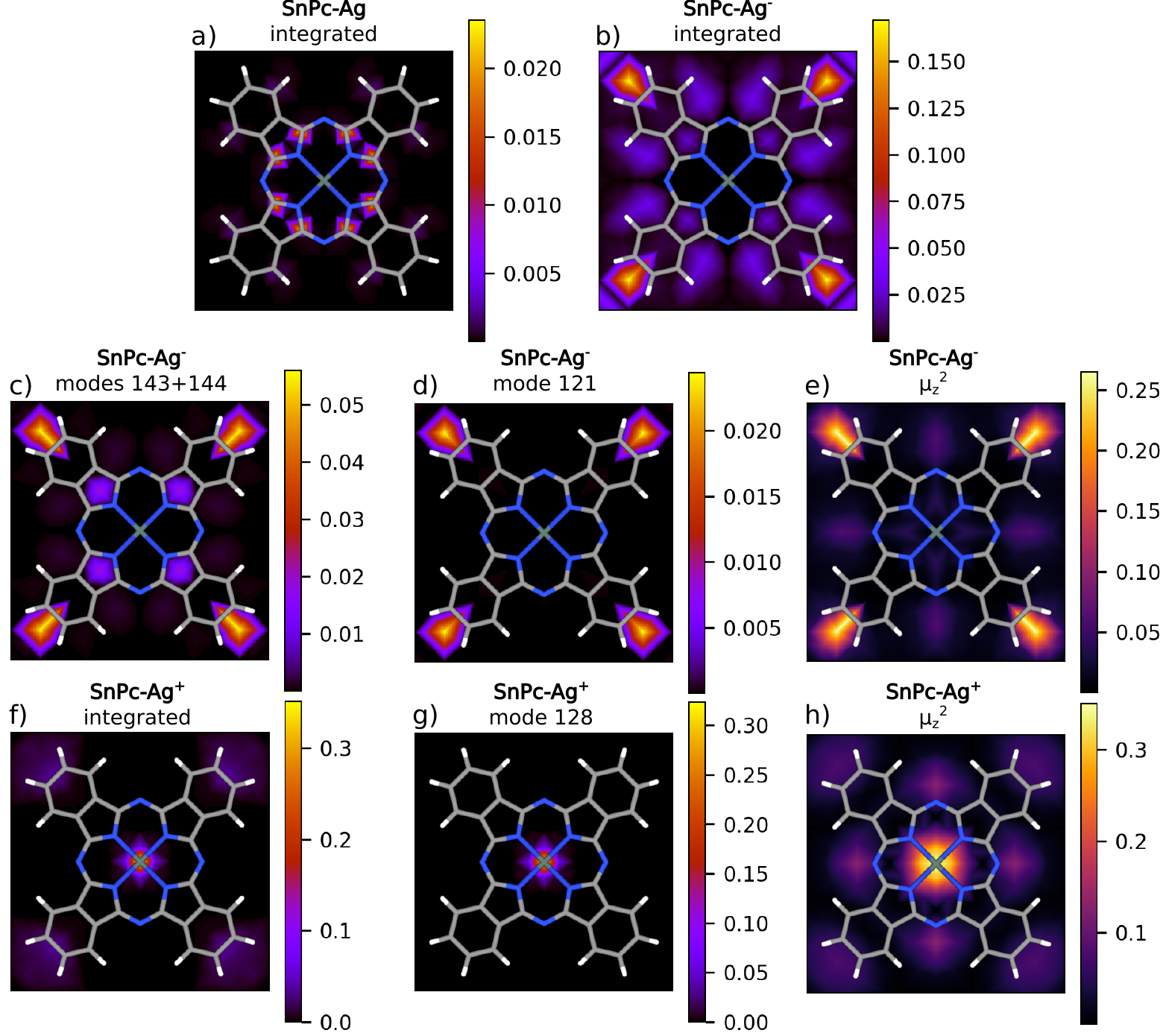}
 \caption{\label{fig:633} Maps for incident radiation at \SI{633}{\nano\meter}.
 a) Integrated intensity map of \textbf{SnPc-Ag}
 b) integrated intensity map,
 c) map of modes 143 and 144,
 d) map of mode 121, and
 e) $\mu_z^2$ of the second $\pi\pi^*$ state in \textbf{SnPc-Ag\textsuperscript{-}} (in a.u.)
 f) integrated intensity map,
 g) map of mode 128, and
 h)$\mu_z^2$ of the second \acrshort{LMCT} state in \textbf{SnPc-Ag\textsuperscript{+}} (in a.u.).
 }
\end{figure*}

\begin{figure}
 \includegraphics{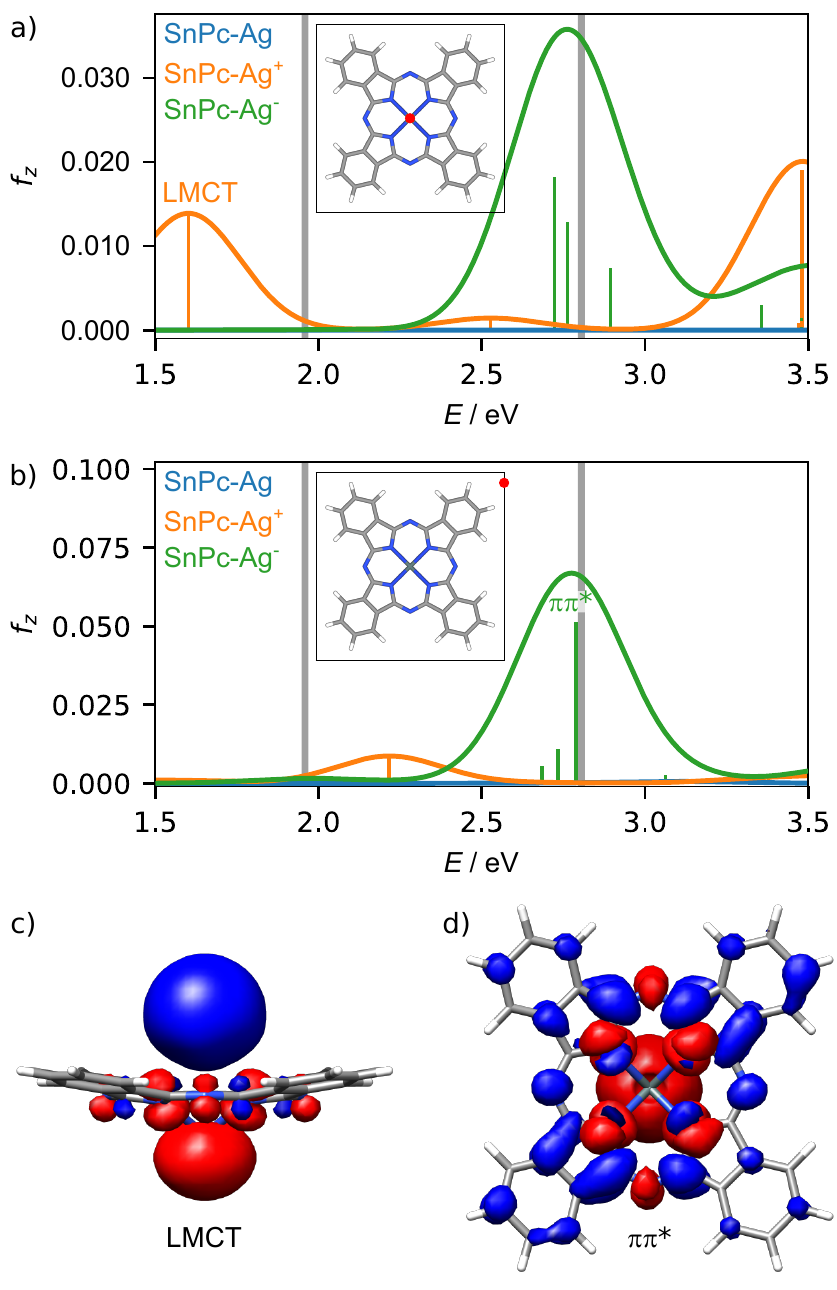}
 \caption{\label{fig:uvvis} a) and b) $z$-polarized UV/vis spectra of \textbf{SnPc-Ag}, \textbf{SnPc-Ag\textsuperscript{+}} and \textbf{SnPc-Ag\textsuperscript{-}} at selected tip positions, see insets. These positions, a) and b), feature the maximum \acrshort{TERS} intensity of \textbf{SnPc-Ag\textsuperscript{+}} at \SI{633}{\nano\meter} and \textbf{SnPc-Ag\textsuperscript{-}} at \SI{442}{\nano\meter} excitation, respectively, see Fig.~\ref{fig:633}f) and Fig.~\ref{fig:442}b); excitation energies (1.96 and \SI{2.81}{\electronvolt}, 633 and \SI{442}{\nano\meter}) are indicated by grey vertical lines.
 Charge density differences for the labelled states:
 c) \acrfull{LMCT} with partial contribution of the Sn atom's $5s$ orbital (side view) and 
 d) $\pi\pi^*$ (top view).}
\end{figure}

While in case of the uncharged \textbf{SnPc-Ag}, tip positions above the pyrrole C atoms stand out at \SI{633}{\nano\meter} excitation (Fig.~\ref{fig:633}a)), 
for the \textbf{SnPc-Ag\textsuperscript{+}} hybrid system, the central point of the molecule stands out in intensity but this time about 4 orders of magnitude brighter than in the non-resonant \acrshort{TERS} map  (recall Fig.~\ref{fig:non-res}d)) with \SI{0.35}{\arbu}, see Fig.~\ref{fig:633}f).
This is due to an \acrshort{LMCT} state representing an electron transfer from Ag to (mainly) Sn (see Fig.~\ref{fig:uvvis}a) and c)) with a strong $\mu_z$ component and with an excitation energy of $\approx \SI{1.5}{\electronvolt}$ -- in (partial) resonance with the exciting light source (see Fig.~\ref{fig:633}h)).
The $z$-component of the transition dipole moment decreases at the neighboring grid points and vanishes quickly for positions away from the center.
Most of the Raman intensity in the center originates from mode 128 ( Fig.~\ref{fig:633}g)), a symmetric stretching of the central $\mathrm{SnN}_4$-fragment, that alone contributes with an intensity of \SI{0.33}{\arbu} at this position.

In case of \textbf{SnPc-Ag\textsuperscript{-}}, four wedge-like structures appear at the molecule's periphery (Fig.~\ref{fig:633}b)).
The second $\pi\pi^*$ state, in resonance at \SI{633}{\nano\meter}, distinctly features a high transition dipole moment in $z$-direction (Fig.~\ref{fig:633}e)) at those bright positions.
This is attributed to the slightly upwards-facing benzene moieties that lead to the otherwise $x,y$ polarized state to bleed into $z$-direction.
Regarding the involved normal modes, the picture is more complex, as several modes contribute to the resonance Raman signal.
The most intense modes are modes~143 (at \SI{0.047}{\arbu}) and 144 (\SI{0.026}{\arbu},  see their sum in Fig.~\ref{fig:633}c)), being stretching vibrations of the bridging C-N bonds, and mode~121 ( Fig.~\ref{fig:633}d)) with \SI{0.025}{\arbu}, a deformation of the pyrrole rings.

\begin{figure*}
 \includegraphics{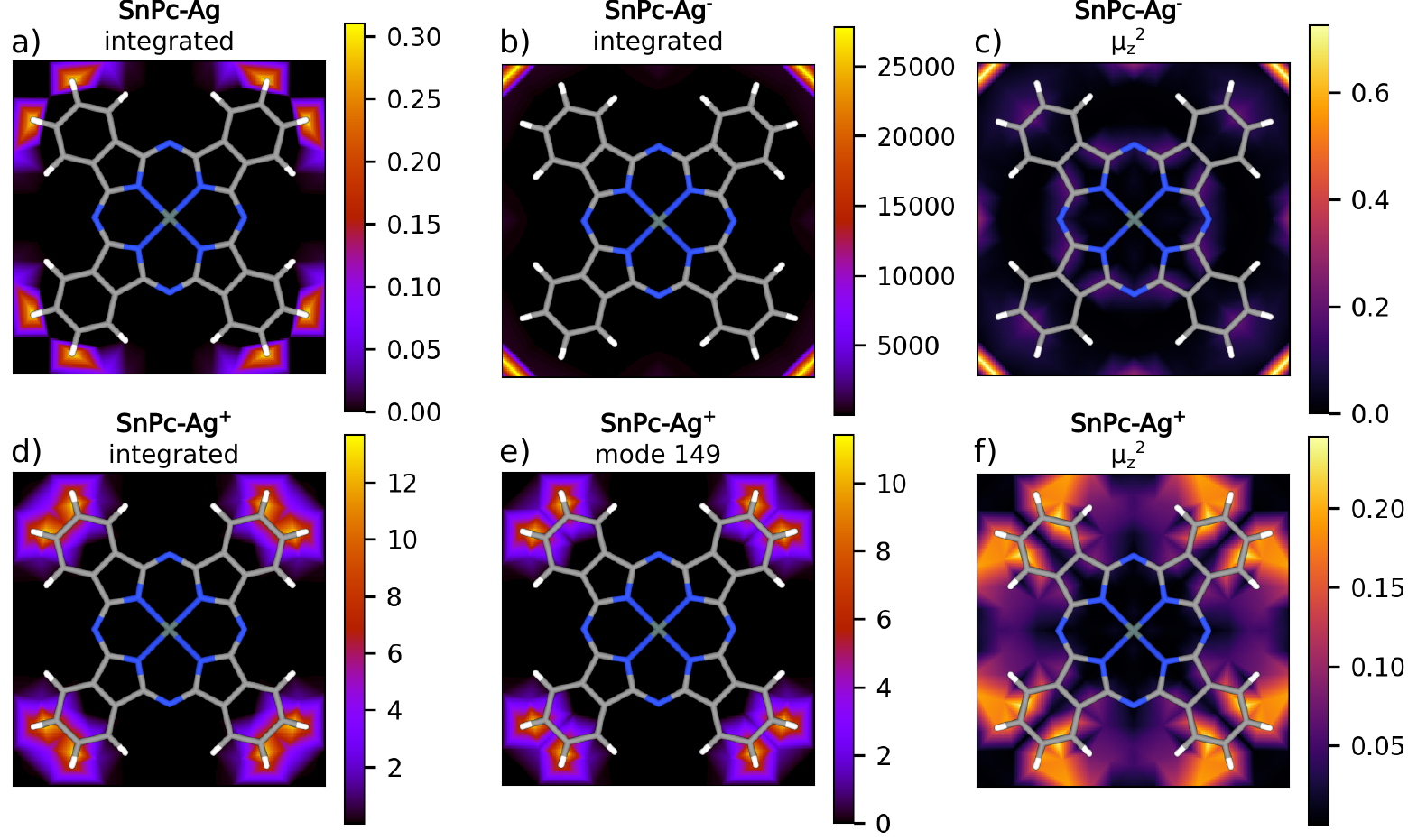}
 \caption{\label{fig:442} Maps for incident radiation at \SI{442}{\nano\meter}.
 a) Integrated intensity map of \textbf{SnPc-Ag}
 b) integrated intensity map and
 c) $\mu_z^2$ of the fourth $\pi\pi^*$ state in \textbf{SnPc-Ag\textsuperscript{-}} (in a.u.)
 d) integrated intensity map, 
  e) map of mode 149, and
 f) $\mu_z^2$ of the fifth \acrshort{LMCT} state in \textbf{SnPc-Ag\textsuperscript{+}} (in a.u.).}
\end{figure*}

At \SI{442}{\nano\meter}, the Raman intensity maximum for both the \textbf{SnPc-Ag} ( Fig.~\ref{fig:442}a)) and the \textbf{SnPc-Ag\textsuperscript{+}} (Fig.~\ref{fig:442}d)) system is localised over the outer $\mathrm{C}_6$ rings.
For \textbf{SnPc-Ag\textsuperscript{+}} there is once more a \acrshort{LMCT} transition with particularly high transition dipole moment in $z$-direction (Fig.~\ref{fig:442}f)) in resonance with the incident light at this position, and again, a single mode (mode~149 at \SI{11.7}{\arbu}) is responsible for most of the overall intensity of \SI{14}{\arbu} (Fig.~\ref{fig:442}e)).
This mode represents a ring deformation of the $\mathrm{C}_6$ ring(s), particularly the one directly below the tip.

As in all discussed scenarios, introducing a charge on the frontmost tip atom significantly alters and enhances the overall intensity of the system's resonance Raman response.
However, in case of the negatively charged tip in \textbf{SnPc-Ag\textsuperscript{-}} and excitation at \SI{442}{\nano\meter}, the signal enhancement is orders of magnitude higher, especially at the periphery of the molecule in the form of a bright band-like structure in Fig.~\ref{fig:442}b).
Comparable to the system under \SI{633}{\nano\meter} illumination, a $\pi\pi^*$ state (see Fig.~\ref{fig:uvvis}b)) is in resonance and its typical polarization in the molecular plane is visible in $z$-direction due to a particlularly large $\mu_z$ (Fig.~\ref{fig:442}c)) at those positions because of the slight upwards tilt of the benzene rings.
The $\pi\pi^*$ state features a distinctly high dimensionless displacement, $\Delta_{e,l}$ (recall Eq.\ref{eq:alpha}), as a result of its pronounced excited state gradient within the Franck-Condon point, along a small number of normal modes across the spectrum. 
All of these modes show similar intensity maps as the summed map in Fig.~\ref{fig:442}b).

\section{Conclusion}

In this contribution, we extend our computational protocol that is able to simulate all three contributions -- non-resonant, resonant and charge transfer -- of the chemical effect in a full quantum chemical approach to also include a static charge on the plasmonic tip.
Thereby, the introduced positive or negative charge reflects the impact on the charge distribution, as predicted for various tip geometries, on the tip's frontmost atom chemically interacting with the sample. 
We investigated a sample molecule, \acrfull{SnPc}, by scanning it with a single, charged silver atom as a tip mimic and observed its \acrshort{TERS} response under different, resonant and non-resonant, conditions to further investigate the source of the proposed sub-molecular resolution.
The system was described by DFT and TDDFT and the Raman response was obtained in $z$-direction; however in principle, the approach can be generalized to other levels of theory and include other illumination-observation geometries.

Under non-resonant irradiation, the hybrid system already reacts strongly to the introduced charge as the overall signal intensity increases by two orders of magnitude.
While for the previously studied, uncharged hybrid system either C-H bending or C-N stretching modes dominate the spectra for different tip positions, \cite{Fiederling_Nanoscale_2020} C-H modes lose importance for the charged systems.
For a positive tip, this leads to a single vibration that dominates the intensity map and is able to accurately pinpoint tip positions over the center of the molecule.
For a negatively charged tip, on the other hand, mostly positions over the extended $\pi$ system of the isoindole moieties provide the highest intensities, this time for several modes involving mainly vibrations of the inner C-N ring.

The general increase in signal intensity after introduction of the charge is observed for resonant excitation as well.
The excited state landscape is changed drastically, however, the molecule's $\pi\pi^*$ states mostly remain unaffected, as several new charge transfer states are introduced.
A single molecule-to-tip charge transfer state dominates in the positively charged hybrid system for both studied wavelengths. 
For low-energy excitation either again positions over the molecule's center are highlighted (as the charge is transferred primarily from the central Sn atom), or, in the high-energy regime, tip positions over the outwards-facing parts of the benzene rings.

The strongest resonant \acrshort{TERS} response is achieved in case of a negative charge on the tip atom.
Here, albeit charge transfer states appear as numerously as in the positive case with opposite directionality, $\pi\pi^*$ states are responsible for the tremendous increase in signal.
While they only played minor roles in the uncharged and positive cases, their transition dipole moment gains a significant component in $z$-direction ($\mu_z$), especially over the slightly upwards tilted outer benzene rings.

Generally, it could be shown that the Raman response of a molecular-plasmonic hybrid system is highly dependent on the tip's position as well as on its charge. 
In addition, the introduction of both a negative or positive charge at the tip's apex alters the region of strongest chemical interactions among the tip mimic and the molecule. In case of the present hybrid systems, this was observed particularly for \textbf{SnPc-Ag\textsuperscript{+}} -- favoring much shorter bonding distances and considerably higher binding energies in comparison to \textbf{SnPc-Ag}.

The incorporation of the electromagnetic effect in the form of more complex fields is the point of interest of ongoing studies, under non-resonant and resonant conditions, as is the addition of the immobilizing surface.

\section*{Supplementary Material}
See supplementary material for excitation energy and transition dipole moment maps of selected $\pi\pi^*$, ligand-to-metal charge transfer and metal-to-ligand charge transfer states as well as potential energies curves and normalized Raman intensities along the $z$-coordinate for the charged and uncharged hybrid systems

\section*{acknowledgements}
The authors gratefully acknowledge funding from the European Research Council (ERC) under the European's Horizon 2020 research and innovation programme 
  -- QUEM-CHEM (grant number 772676), ``Time- and space-resolved ultrafast dynamics in molecular plasmonic hybrid systems``.
  This work was further funded by the Deutsche Forschungsgemeinschaft (DFG, German Research Foundation) -- project number A4  -- SFB 1375.
  All calculations were performed at the Uni\-ver\-si\-täts\-re\-chen\-zen\-trum of the Friedrich Schiller University Jena.

\section*{AIP publishing data sharing policy}
The data that support the findings of this study are available from the corresponding author upon reasonable request.

%

\end{document}